# Anomalous Tail Effect on Resistivity Transition and Weak-link Behavior of Iron Based Superconductor

Y. Sun[1], Y. Ding[1], J. C. Zhuang[1], L. J. Cui[1], X. P. Yuan[1], Z. X. Shi[1*], Z. A. Ren[2]

1. Department of Physics, Southeast University, Nanjing, China 211189

2. Institute of Physics and Beijing National Laboratory for Condensed Matter Physics, Chinese Academy of Sciences, Beijing 100190, P. R. China

Email: zxshi@seu.edu.cn

**Abstract**

Temperature dependent resistivity of the iron-based superconductor $NdFeAsO_{0.88}F_{0.12}$ was measured under different applied fields and excitation currents. Arrhenius plot shows an anomalous tail effect, which contains obvious two resistivity dropping stages. The first is caused by the normal superconducting transition, and the second is supposed to be related to the weak-link between the grains. A model for the resistivity dropping related to the weak-link behavior is proposed, which is based on the Josephson junctions formed by the impurities in grain boundaries like FeAs, $Sm_2O_3$ and cracks together with the adjacent grains. These Josephson junctions can be easily broken by the applied fields and the excitations currents, leading to the anomalous resistivity tail in many polycrystalline iron-based superconductors. The calculated resistivity dropping agrees well with the experimental data, which manifests the correctness of the explanation of the obtained anomalous tail effect.

## 1. Introduction

The recent discovery of superconductivity at 26 K in the iron oxypnictide LaFeAs(O, F) [1] has stimulated great interests among the condensed-matter physics community. Tremendous work has been carried out, leading to the emergence of novel iron-based superconductor families with different crystal structures: 1111 (REFeAs(O,F)), 122 ($(Ba,K)Fe_2As_2$) [2], 111 (LiFeAs) [3] and 11 (Fe(Se,Te)) [4]. The transition temperature $T_c$ of REFeAs(O, F) superconductors is over 50 K when La is replaced by Sm [5], Gd [6] or Tb [7], and the upper critical field $H_{c2}$ (0 K) exceeds 100 T [8], suggesting promising potential applications. However, these materials are significantly anisotropic due to the layered structure, and have a low carrier density in the order of $10^{21}$ $cm^{-3}$ [9]. High $H_{c2}$ also implies a very short coherence length $\xi$. The similarity of these basic superconducting properties to the cuprate-based superconductors (CBS) indicates the possible existence of weak links and electromagnetically granular behavior, as already investigated by magnetization measurements and magneto-optic imaging in previous reports [10] [11] [12] [13]. Furthermore, almost all the investigations of the critical current density ($J_c$) of 1111 polycrystalline samples have shown significant evidence for granularity and low intergranular $J_c$ values [14]. Although the weak link behavior was testified in many experiment results, its origin and the mechanism under it are still unknown. This problem is crucial to understand whether the electromagnetically granular behavior is the intrinsic properties like the case for cuprates or just caused by the preparation process. And the weak link behavior will also complicate the study of the vortex dynamics. This can be witnessed from the different experimental results on the effective pinning barrier of the superconducting $NdFeAsO_{1-x}F_x$ by different groups [15] [16].

In this work, temperature dependent resistivity transition measured under different applied fields and excitation currents shows an anomalous tail effect. This anomalous tail effect have also been seen in previous report on iron-based superconductor [17] [18] but without being paid enough attention. In this article, we detailed studied the origin as well as the mechanism of the anomalous resistivity tail, and proposed a model to calculate this effect.

## 2. Experimental Results

The superconducting $NdFeAsO_{0.88}F_{0.12}$ was prepared by the two-step solid-state reaction method [19]. The stoichiometric mixture of starting materials was ground thoroughly and pressed into pellets, then sintered in an evacuated quartz tube at 1150-1165 $^{\circ}C$ for 60-72 h. The crystal structure of the sample was characterized by powder X-ray diffraction (XRD) on an MXP18A-HF–type diffractometer with Cu-Kα radiation from 20° to 80° with a step of 0.01°. For the transport measurements, the sample was cut into a rectangular shape with dimension of 5.4 mm (length) × 3.05 mm (width) × 0.76 mm (thickness). The standard four-probe technique under magnetic field up to 9 T was performed on a physical property measurement system (PPMS, Quantum Design).

Figure 1 shows the XRD pattern of $NdFeAsO_{0.88}F_{0.12}$. Almost all the diffraction peaks can be well indexed on the basis of tetragonal ZrCuSiAs-type structure with a P4/nmm space group, confirming the main phases are $NdFeAsO_{1-x}F_x$. However, there still exists some small amount of impurities which were detected by the EDX in our previous report [20]. As reported by F. Kametani [21], non-superconducting Fe–As and $RE_2O_3$ occupy at least three quarters of the RE1111 grain boundaries (GBs), blocking the transport supercurrent. This effect maybe one of the main reasons of the strongly field-dependent global current density $J_c$ and the large difference between the obtained intergrain critical current density and the intragrain current density [13].

To investigate the origin as well as the effect of the weak-link behavior, temperature dependence of the resistivity with applied fields up to 9 T was measured and plotted in Figure. 2 (a). The sample shows a superconducting transition temperature of about 44 K. Below $T_c$, the resistivity drops quickly, and is significantly broadened by the increasing magnetic fields. Then the resistivity displays a long tail until it merges into the zero resistivity flat floor. In order to clearly observe the superconducting transition behavior, the resistivity of $NdFeAsO_{0.88}F_{0.12}$ was redrawn in the Arrhenius plot, which was shown in Figure. 2 (b). An anomalous resistivity tail was witnessed, which undergoes two distinct dropping stages. At first, the resistivity decreases quickly, and is broadened by the magnetic field. This is common to see in high-$T_c$ cuprates and oxypnictides, which is attributed to the strong thermal fluctuations that from the high transition temperature, short coherence length, large anisotropy, as well as the inhomogeneity of the polycrystalline sample. Then the resistivity decreases more slowly and even cannot reach the zero resistivity in the measurement limit when the applied field is large enough. The anomalous second resistivity doping stage is supposed to be related to the weak-link behavior of the grains, which is easily broken by the applied fields. By the way, the fluctuation of the data may be caused by the thermal activated vortex motion including the homogeneous vortex flow or creep and the vortex flow along some of the inter-grain boundaries.

If the second resistivity dropping stage is caused by the weak-link behavior, it will be sensitive to the current density. Thus Arrhenius plot of the temperature dependent resistivity under zero field with different excitation currents was shown in Figure. 3. As expected, obvious two resistivity dropping stages can also be seen. For the absence of the magnetic field, the field induced resistivity broadening, as well as the fluctuation of the data are not observed. And the latter makes the quantitative analysis of the weak-link behavior possible. We must point out that the real current density applied in our sample is much large than that calculated from the excitation current divided by the cross area. For the non-superconducting impurities like Fe–As and $RE_2O_3$ occupying a large part of the GBs [21], making the active current path certainly much smaller than the geometrical cross-section of the sample. If the active current path can be calculated, the relationship between superconducting critical current density $J_c$ and temperature can be obtained. Unfortunately, as far as we know, there's still no way to settle this problem. Thus we can only get the critical current $I_c$ vs temperature, which was shown in the inset of Figure. 3. The criterion was chosen as 1% normal state resistivity, which was usually used in obtaining the lower critical field $H_{c1}$. This gives out the critical temperatures $T_c^*$ at which almost all the grains become superconducting under different excitation currents. And these temperatures are roughly taken as the boundary of the two resistivity dropping stages. Above the boundary, the resistivity dropping is caused by the superconducting transition. While below the boundary, the resistivity dropping is attributed to the weak-link behavior.

Low temperature laser scanning microscopy and scanning electron microscopy [22] revealed that Fe–As is a normal-metal wetting-phase that surrounds Sm-1111 grains, producing a dense array of superconducting-normal-superconducting contacts, which is in accordance with our experimental results on Nd-1111 sample. And the impurity $RE_2O_3$ and cracks together with their adjacent grains may also produce some superconductor-insulator-superconductor contacts. Furthermore, our results also show that the two adjacent grains in these contacts are coupled. Otherwise, the resistivity dropping in the second stage should manifest a metal behavior, and would not that sensitive to the applied fields and the excitation currents. The coupled contacts can form either the quasiparticle tunneling or the Josephson junction depending on the distance between the two adjacent grains. The former can be eliminated for the hopping resistance is proportional to $\exp(\Delta(T)/k_B T)$ [23-25] leading to an upturn curve in the Arrhenius plot, which is not observed in our experiment results even when the applied fields and the excitation currents are very large. Taking the Josephson junction into consideration, when the temperature decreases to the critical temperature $T_c^*$, almost all the separate grains become superconducting together with parts of the non-superconducting Josephson junctions. The non-superconducting Josephson junctions come from the relatively large current density which exceeds their critical current density $J_c$. Then as the temperature continues to decrease below $T_c^*$, $J_c$ continues to increase, which causes some non-superconducting Josephson junctions turn superconducting. This explains how the resistivity decreases in the second region. Figure. 3 shows that when the excitation current is small, the resistivity drops quickly to the zero resistivity state as the temperature decreases. That's because the relatively small excitation current is easily overcome by the increment of the $J_c$. Then as the excitation current increases, the resistivity drops more slowly, which can be witnessed from the continuous reducing of the curve's slop. At last when the excitation current is large enough that most of the junctions cannot turn superconducting, there exists no superconducting path in the sample, and the resistivity cannot decrease to zero.

Now we turn back to the resistivity measured under different magnetic fields. When the magnetic fields are applied on the sample, the distance between the two adjacent grains will be increased because of the penetration of the applied fields, leading to the decrease of the critical current density [26]. Thus the effect of the applied fields on the Josephson junction is similar to that of the excitation currents. The resistivity of the second stage in Figure. 2 (b) also drops quickly to the zero field state at a small applied field, while cannot turn to totally superconducting phase when the field is large.

## 3. Model and Numerical Simulations

To confirm the correctness of attributing the anomalous tail effect to the weak link behavior, the resistivity transition was calculated based on the explanation above. In order to derive a model for the resistivity dropping in the weak linked region, a quasi one-dimensional current path is taken into consideration first. This path can be seen as a line of Nd1111 grains separated by some non-superconducting impurities and cracks. When temperature reduced to the critical temperature $T_c^*$, all the Nd1111 grains turn to superconducting. Then the path can be seen as a strip of Josephson junctions. For the sake of simplicity, distance between two adjacent grains is assumed to be randomly distributed, thus the magnitude of the critical current density $J_c$ of the grains at a certain temperature are also randomly distributed, for $J_c$ is one-to-one correspondence to the distance between two adjacent grains [27] .When the excitation current density is larger than the critical current densities of all the grains, the total resistance $R$ of the current path is equal to $\sum_{i,j} R_{ij}$, where $R_{ij}$ is the normal resistance between the neighboring grains. As temperature decreases, the critical current density $J_c$ continues to increase accompanied by the decrement of the resistance. Thus resistance at the temperature of $T$ can be expressed as $R(T)=\sum_{i,j} R_{ij}[1-J_c(T)/J_c(0)]$ (1), where the $J_c(0)$ is the critical current density at the temperature of zero. Ambegaokar–Baratoff (AB) formula [27] gives out that $J_c(T)=\frac{\pi\Delta(T)}{2eR_{ij}}th\frac{\Delta(T)}{2k_BT}$, where $k_B$ is the Boltzmann's constant, and $\Delta(T)$ is the superconducting gap, which can be roughly expressed by using the BCS type gap formula $\Delta^{(s)}(T)=\Delta^{(s)}(0)\tanh\{1.82[1.018(\frac{T_c}{T}-1)]^{0.51}\}$ [28]. Taking the expression of $J_c$ into the formula of $R(T)$, we can get the relation that $R(T)=\sum_{i,j} R_{ij}[1-(\Delta(T)/\Delta(0))th(\Delta(T)/2k_BT)]$ (2). As discussed above in the experiment results, the effect of both the applied fields and the excitation currents can be seen as reducing the critical current density, thus slow down the decrement of the resistivity. The influence can be considered by adding a field and current dependent prefactor A ($B$, $I$) into the expression of resistance, which can be rewritten as $R(T)=\sum_{i,j} R_{ij}[1-A(B,I)J_c(T)/J_c(0)]=\sum_{i,j} R_{ij}[1-A(B,I)(\Delta(T)/\Delta(0))th(\Delta(T)/2k_BT)]$ (3). The prefactor $A$ is equal to 1 at the zero field and very small excitation current, decreasing as the increasing of the applied field or excitation current. In order to evaluate this expression numerically, we assume that the resistances between the grains are equal. Thus the resistance of the current path can be calculated basically by counting the number of grains with the critical current density below the excitation current density.

Numerical simulation results of the temperature-dependent resistivity based on the Eq. (3) are shown in Figure. 4 in Arrhenius plot. Different curves in the figure are calculated by different values of $A$. The arrow in the figure indicates the direction of the decreasing of the prefactor $A$, which also manifests the increase of applied field or excitation current as discussed above. The simulation results show that when the prefactor A is close to 1, the resistivity can drop to zero quickly as the temperature decreases. This behavior is similar to that observed in the experiment when the magnetic field and the excitation current are very small. Then the resistivity drops slower as the decrease of the prefactor $A$, which is also in accordance with the experiment results on the increasing of the applied field or excitation current. At last the resistivity cannot reach zero just like the experimental case that the field or the excitation current is large enough, which means there still exists some non-superconducting junctions even at very low temperature. As discussed above, the calculated resistivity dropping agrees well with the experimental data, manifesting the correctness of our explanation on the anomalous tail effect. Our results also give out a new and easier way to determine the existence and the origin of the intergranular weak link.

Our results are calculated based on a single current path without considering the electric short circuit in the resistor networks, which will accelerate the resistivity dropping. The short circuit effect can be ignored when the temperature is relatively large but is really witnessed at very low temperature from the data of 50 mA and 100 mA in Figure. 3, which show an abrupt dropping before turning to zero resistivity. To solve this problem, the current percolation should be introduced into our model, which proved successfully in granular superconductor [29], cuprate superconductor [30], $MgB_2$ [31], and also found in the superconducting $SmFeAsO_{1-x}F_x$ [32]. While the current percolation in iron-based superconductor is still unclear so far, therefore more work should be done to clarify this issue.

## 4. Conclusion

In summary, we systematically studied the temperature dependent resistivity of the iron-based superconductor $NdFeAsO_{0.88}F_{0.12}$ under different applied fields and excitation currents. An anomalous tail effect with two resistivity dropping stages was found and attributed to the intergranular weak-link behavior. Then the origin and the mechanism of the weak-link behavior are explained by the formation of Josephson junctions by the non-superconducting impurities and cracks together with the adjacent grains. A model based on this explanation is proposed, and the calculation results agree well with the experimental data.

## 5. Acknowledgments

We are very grateful to Professors Haihu Wen and Yanwei Ma for their help. This work was supported by the Natural Science Foundation of China, the Ministry of Science and Technology of China (973 project: No. 2011CBA00100), National Science Foundation of Jiangsu Province of China (Grant No. BK2010421).

# Figure captions

Figure 1: X-ray powder diffraction patterns of superconducting $NdFeAsO_{0.88}F_{0.12}$.

Figure 2: Resistivity transition of $NdFeAsO_{0.88}F_{0.12}$ in ρ-T (a) and Arrhenius plot (b) with increasing magnetic field up to 9 T.

Figure 3: Resistivity transition of $NdFeAsO_{0.88}F_{0.12}$ in Arrhenius plot with increasing excitation current. Inset plots the critical current $I_c$ vs the critical temperature $T_c^*$ at which almost all the grains have turned to superconducting phase under different excitation currents

Figure 4: Numerical simulations of the temperature dependence of resistivity related to the weak link behavior. The arrow indicates the direction of increase of magnetic field or excitation current.

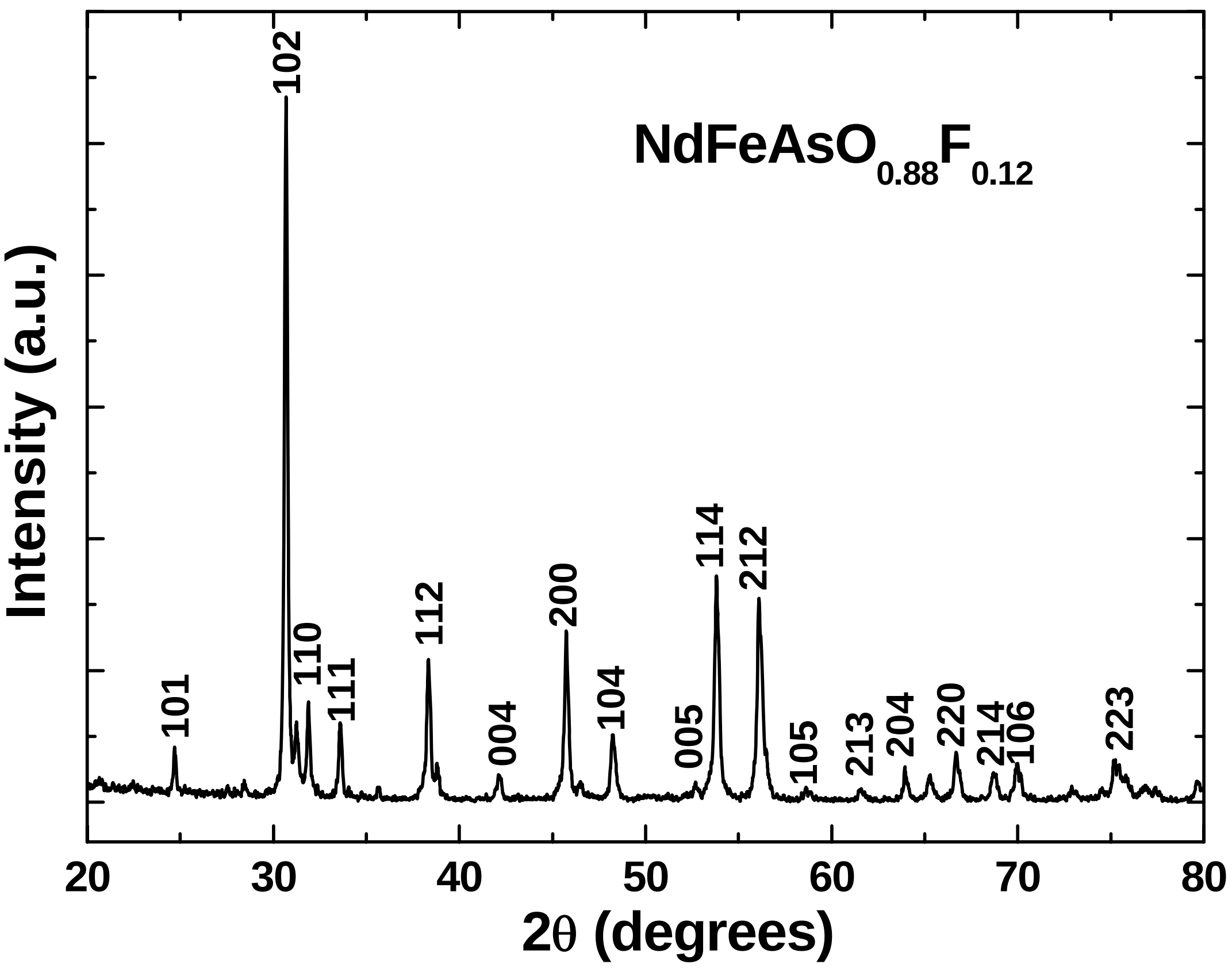

**Figure. 1 Y. Sun *et al.***

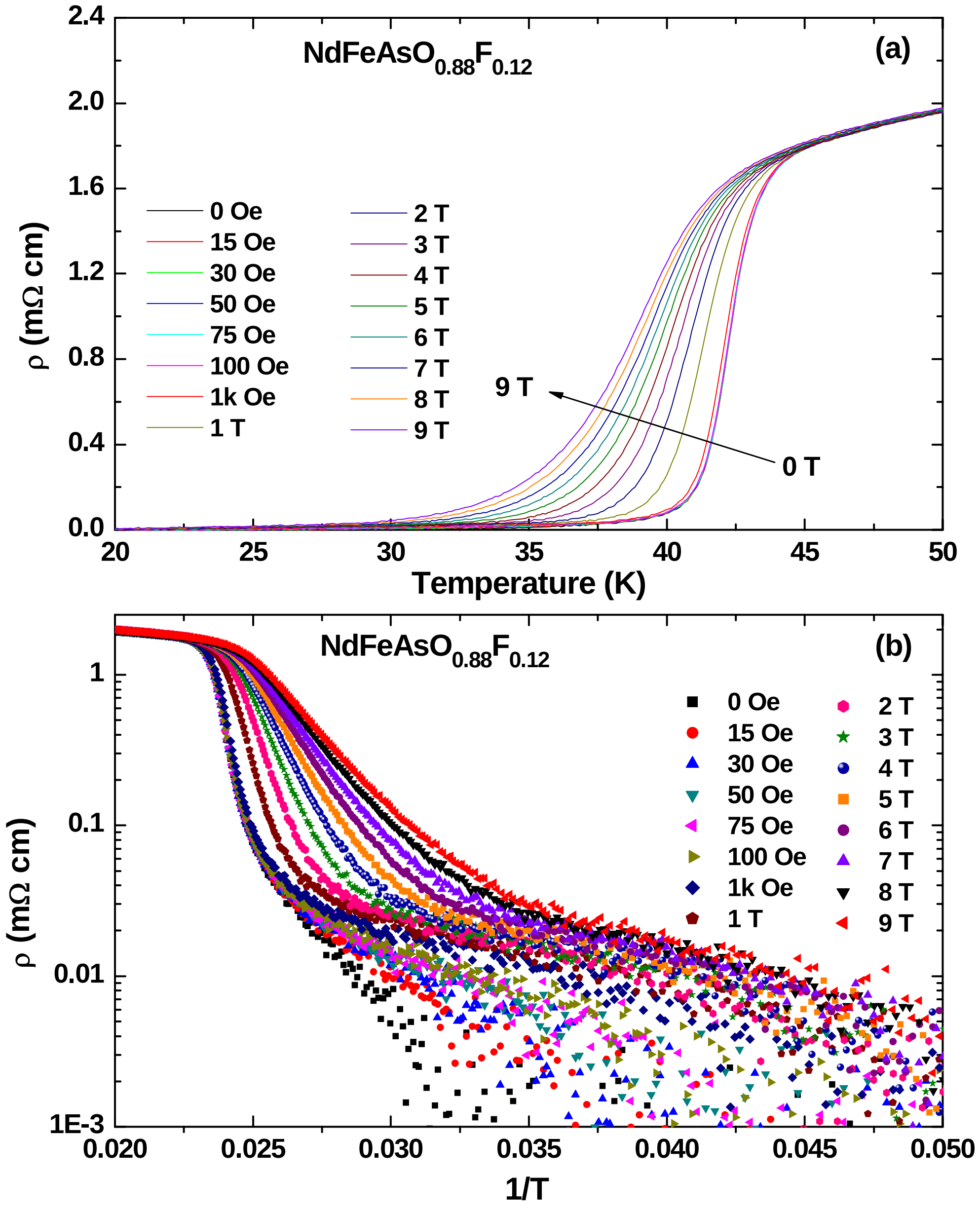

**Figure. 2 Y. Sun *et al.***

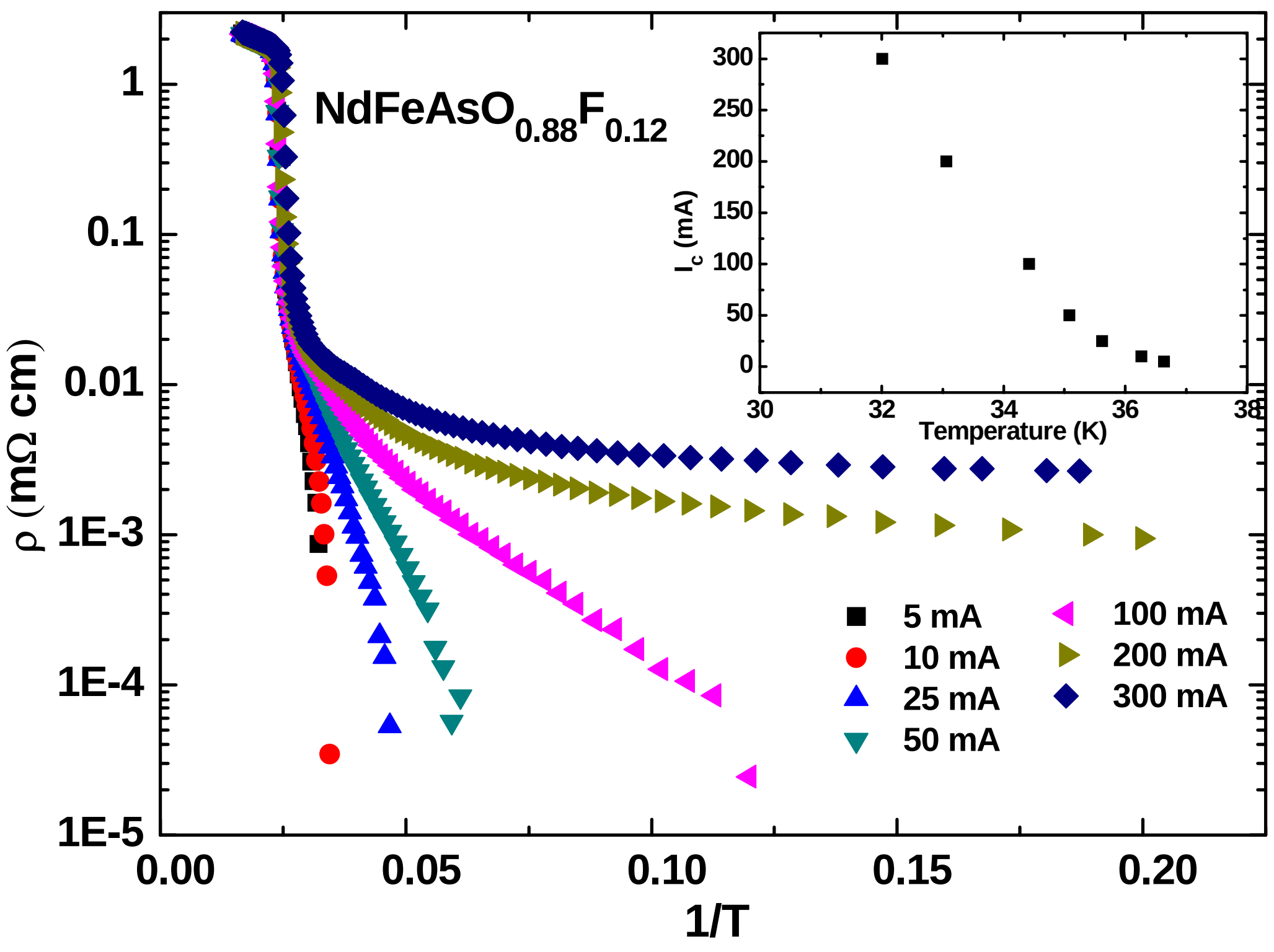

**Figure. 3 Y. Sun *et al.***

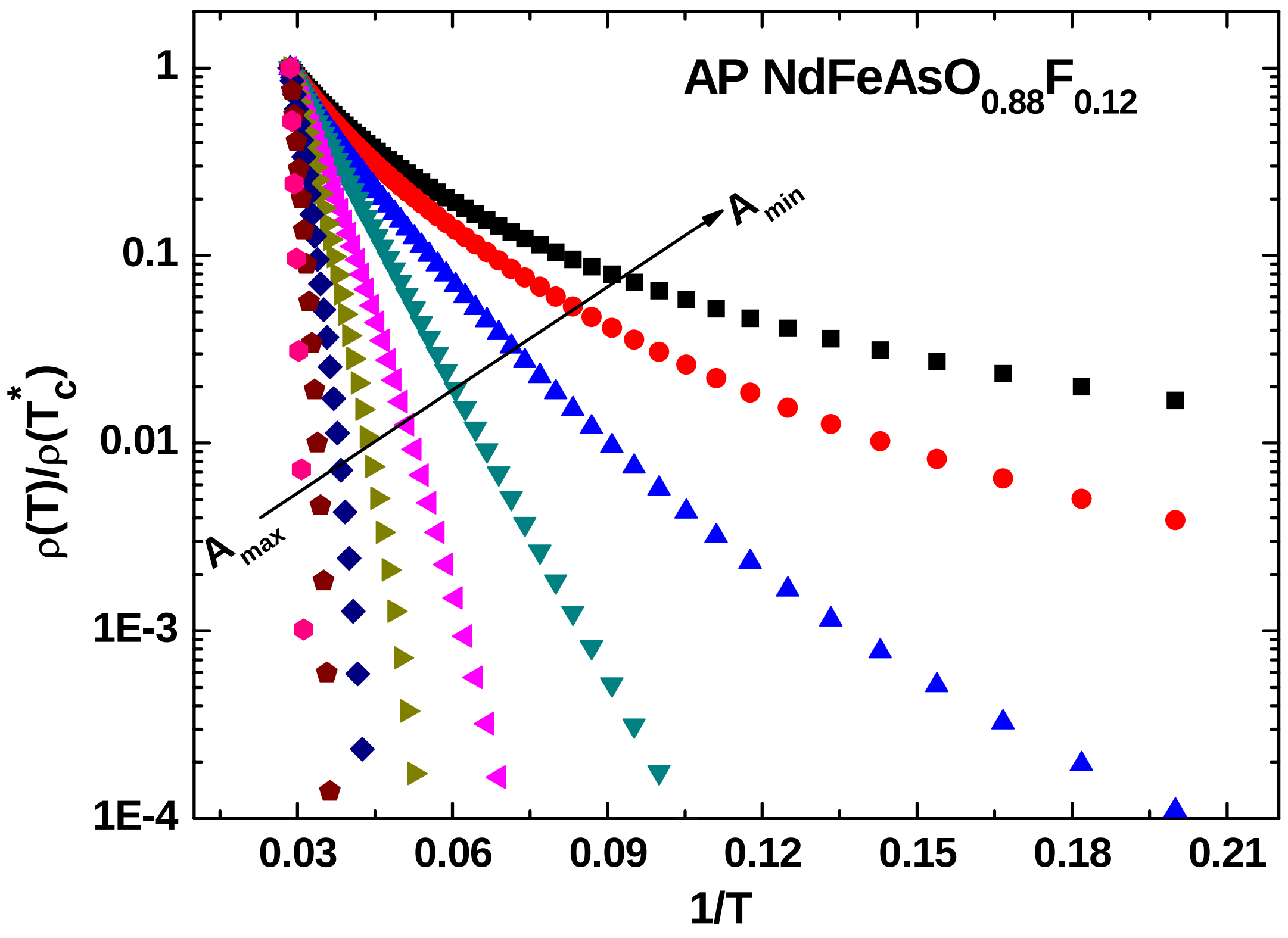

**Figure. 4 Y. Sun *et al.***